\documentstyle[epsf]{article}
\begin{document}

\begin{center}

{\Large {\bf 
               Evolution of close binaries after 
               the  burst of starformation for different
               IMFs

\bigskip
\bigskip

S.B. Popov, \\
}}

{{\large \bf (http://xray.sai.msi.su/\~ \, polar)}}\\

{\Large {\bf

V.M. Lipunov, M.E. Prokhorov \& K.A. Postnov\\

}}

\bigskip

{{\large \bf
Moscow State University\\

Sternberg Astronomical Institute
}}

\end{center}

\bigskip

{\Large

\centerline{{\bf Abstract}}
}

\bigskip

{\large

    We use "Scenario Machine" -- the population synthesis
simulator -- to calculate the evolution of  populations of selected
types of X-ray sources after a starformation burst with the total mass
in binaries $(1$--$1.5) \cdot 10^6 M_{\odot}$ 
during the first 20 Myr after a burst.
We present here the  results of  two sets of runs of the program.

In the first set we examined the following types of close binaries:
transient sources--  neutron stars with  Be- stars; ``X-ray pulsars''--
 neutron stars in pairs with  supergiants; Cyg X-1-like sources--
 black holes with  supergiants; ``SS443-like sources''--  superaccreting
black holes. We used two values of the exponent $\alpha$ in the initial
mass function: 2.35 (Salpeter's function) and 1 (``flat spectrum'').
The calculations were made for the folowing values of the upper limit of the
mass function: 120 and 30 $M_{\odot}$.
For the ``flat spectrum'', suggested in (Contini et al, 1995), 
the number of sources  of all types  significantly
increased. With the ``flat spectrum''
and with the upper mass limit 120 $M_{\odot}$ we obtained 
hundreds of sources of {\it all} calculated types.
Decreasing of the upper mass limit below the critical mass
of a black hole formation increase the number of transient sources   
with neutron stars up to $\approx 300$.

In the second set we examined the evolution of 
of  12 other  types of  X-ray sources  for $\alpha=1$, $\alpha=1.35$ 
and $\alpha=2.35$ and
for three upper mass limits: 120 $M_{\odot}$, 60 $M_{\odot}$, 40 $M_{\odot}$
(see Perez-Olia \& Colina 1995 for the reasons for such upper limits)
on the same time scale 20 Myr after a star formation burst.

}

\bigskip


\section{Introduction. Why do we do it?}


 Theory of stellar evolution and one of the strongest tools 
of that theory -- population synthesis -- 
are now quickly developing branches of astrophysics.
Very often  only the evolution of single stars is modelled.
But it is well known
that about 50\% of all stars are members of  binary systems,
and a lot of different astrophysical objects are products
of the evolution of binary stars. We argue, that often it is
necessary to take into account the evolution of close binaries
while using the population synthesis in order to avoid serious  errors. 

 Partly this work was stimulated by the article by Contini et al. (1995),
where the authors suggested a very unusial form of the initial mass function
(IMF) for the explanation of the observed properties
of the  galaxy Mrk 712 . They suggested the ``flat'' IMF with the exponent  
$\alpha=1$ instead of the Salpeter's value  $\alpha=2.35$.
Contini et al. (1995) didn't take into account binary systems, so
no words about the influence of such IMF  
on the populations of close binary stars could be said.
Later Shaerer (1996) showed that the observations could be explained
without the IMF with $\alpha=1$. 
Here we try to determine the  influence of the 
variations of the IMF on the evolution of compact binaries.

Previously (Lipunov et al, 1996a) we used  the
``Scenario Machine'' for  calculations of  populations of
X-- ray sources at the Galactic center. Here we model a general
situation --- we made calculations for a typical starformation burst.
We present two sets of calculations. In the first one only four types
of binary sources were calculated for two values of the upper mass limit
for two values of $\alpha$. In the second one we show results on 
twelve types of binary sources with significant X-ray luminosity for three
values of the upper mass limit for three values of $\alpha$.  



\section{Model. How do we do it?}


Monte-Carlo method for statistical simulation of binary evolution
was originally proposed by Kornilov \& Lipunov (1983a,b) for massive
binaries
and developed later by Lipunov \& Postnov (1987) for low-massive binaries.
Dewey \& Cordes (1987) applied an analogous method
for analysis of radio pulsar statistics, and de Kool (1992)
investigated  by the Monte-Carlo method  the formation  of the galactic
cataclysmic variables. 

Monte-Carlo simulations of binary star evolution allows one to
investigate the evolution of a large ensemble of binaries  and to
estimate the number of binaries at different
evolutionary stages. Inevitable simplifications in the
analytical description of the binary evolution that we allow in our
extensive numerical calculations, make those numbers
approximate to a factor of 2-3.  However, the inaccuracy of direct
calculations  giving the numbers of different binary types
in the Galaxy (see e.g. Iben \& Tutukov 1984, 
van den Heuvel 1994) seems to be comparable to what follows from the
simplifications in the binary evolution treatment.  

In our analysis of binary evolution, we use the ``Scenario Machine'', a
computer code that incorporates all current scenarios of binary
evolution 
and takes into account the influence of magnetic field of
compact objects on their observational appearance. A detailed description
of the computational techniques and input assumptions is summarized
elsewhere (Lipunov et al. 1996b), and here we briefly list only principal
parameters and initial distributions.

We trace the evolution of binary systems during the first 20 Myr after
their
formation in a starformation burst. Obviously, only
massive enough stars (with masses $\ge 8-10~ {\rm M}_\odot$) can evolve off
the main sequence during the time as short as this to yield compact
remnants (NSs and BHs).
Therefore we consider only massive binaries, i.e. those having the mass of
the primary (more massive) component in the range of $10-120$
${\rm M}_\odot$.

The distribution in orbital separations is taken as deduced from
observations:
\begin{equation}
f(\log a) ={\rm const}\,,\qquad \max~\{10~ {\rm R}_\odot,~\hbox{Roche
Lobe}~M
(M_1)\} < \log a < 10^4~{\rm R}_\odot.
\end{equation}

We assume that a NS with a mass of $1.4~ {\rm M}_{\odot}$ is formed
as result of the collapse of a star, whose core mass prior to collapse was
$M_*\sim (2.5-35)~{\rm M}_{\odot}$. This corresponds to an initial mass 
range $\sim (10 - 60)~{\rm M}_{\odot}$, taking into account that a massive
star can lose more than $\sim (10-20)\%$ of its initial mass during the   
evolution with a strong stellar wind.

The most massive stars are assumed to collapse into a BH once
their mass before the collapse is $M>M_{cr}=35~ {\rm M}_\odot$ (which
would correspond to an initial mass of the ZAMS star as high as $\sim
60~ {\rm M}_\odot$ since a substantial mass loss due to a strong
stellar wind occurs for the most massive stars).  The BH mass is
calculated as $M_{bh}=k_{bh}M_{cr}$, where the parameter $k_{bh}$ is
taken to be 0.7.

The mass limit for NS (the Oppenheimer-Volkoff limit) is taken to be
$M_{OV}=2.5~ {\rm M}_\odot$, which corresponds to a hard equation of
state of the NS matter.

We made calculations for several values of the coefficient $\alpha$:

\begin{equation}
   \frac{dN}{dM} \propto M^{-\alpha}
\end{equation}

We calculated $10^7$ systems in every run of the program.
For the normalization we used the lower mass limit  $0.1 M_{\odot}$.
Then the results were normalized to the total mass of binary stars 
in the starformation burst.
We also used different values of the upper mass limit.

We also take into account that the collapse of a massive star into a NS
can be asymmetrical, so that
an additional kick velocity, $v_{kick}$, presumably randomly oriented in
space, should be imparted to the newborn NS.
We used the velocity distribution in the form
obtained by Lyne \& Lorimer (1994) with the characteristic value 200 km/s
(twice less than in Lyne \& Lorimer (1994)).



\newpage

\section{Results. What have we done?}


 On the figures we show the results of our calculations for both sets.
On all graphs on the X- axis we show the time after the 
starformation burst in Myrs, on the Y- axis --- number of 
the  sources of the selected type that exist at the particular moment
(not the birth rate of the sources!).

On figure 1 we present the results of our calculations of the
evolution of populations of X- ray sources of the four types (the first set)
for the upper mass limit $ 120\, M_{\odot}$ (upper graph) and
$30\, M_{\odot}$ (lower graph).

\begin{itemize}

\item
Transient sources- a neutron star with a Be- star (graphs (1a)
and (2a)).

\item 
``X-ray pulsars''-- a neutron star in pair with a supergiant 
(of course not all X- ray pulsars belong to this type of sources,
but all systems of that type should appear as X- ray pulsars)
(graphs (2b)and (1b)).

\item 
Black holes with  supergiants. Cyg X-1 is a prototype of the sources
of that kind (graph (2d)).

\item 
And at last superaccreting black holes (graphs (2c) and (1c)).  
We call this type -- ``SS 433''-like
sources, as the well known object SS 433 can belong to that class
of astrophysical sources.

\end{itemize}

Solid line --- Salpeter's mass- function, dot-dashed line --- 
``flat'' IMF. The calculated numbers were normalized for $1.5\cdot 10^6
\, M_{\odot}$ in binary stars.

\bigskip

On figures 2-4 we show our calculations for X- ray sources of 12
differnt types (the second set, see a brief description of that types below).

\begin{itemize}

\item
Figure 2 --- $\alpha=1$, 

\item
Figure 3 --- $\alpha=1.35$,

\item
Figure 4 --- $\alpha=2.35$. 

\end{itemize}

\noindent
For upper mass limits:

\begin{itemize}

\item
$120 M_{\odot}$ -- solid lines,

\item
$60 M_{\odot}$ -- dashed lines,

\item
$40 M_{\odot}$ -- dotted lines.

\end{itemize}

 The calculated numbers were normalized for $1\cdot 10^6 
\, M_{\odot}$ in binary stars. We show on the figures 2-4
only systems with X-ray luminosity greater than $10^{33}\, erg/s$.

Curves were not smoothed. We calculated $10^7$ binary systems in every run,
and then the results were normalized.

We used the  ``flat'' mass ratio function, i.e. binary systems with
any mass ratio appear with the same probability. The results can be 
renormalized to any other form of the mass ratio function.


\begin{figure}
\epsfxsize=\hsize
\centerline{{\epsfbox{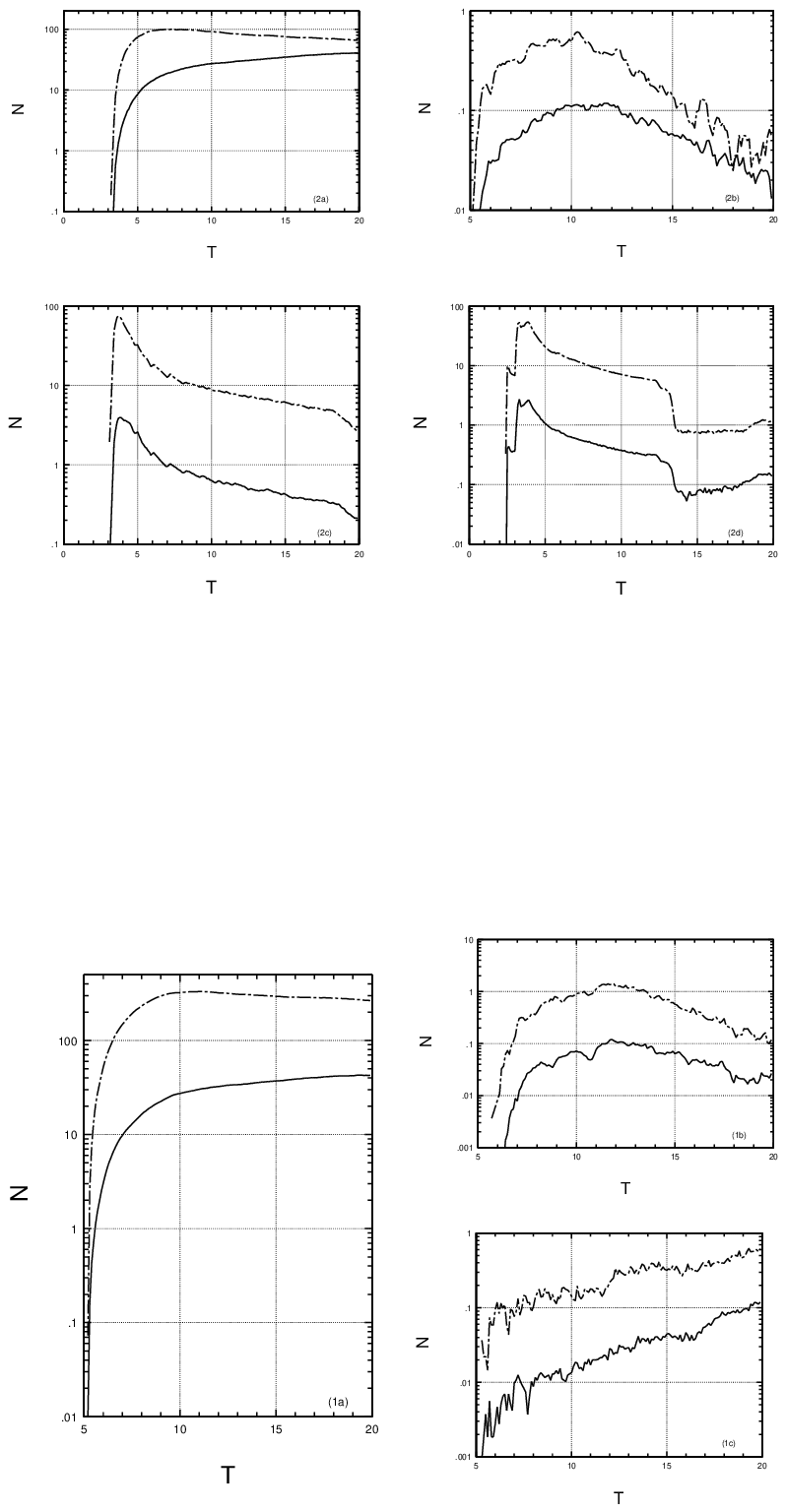}}}    
\end{figure}

\begin{figure}
\epsfxsize=\hsize
\centerline{{\epsfbox{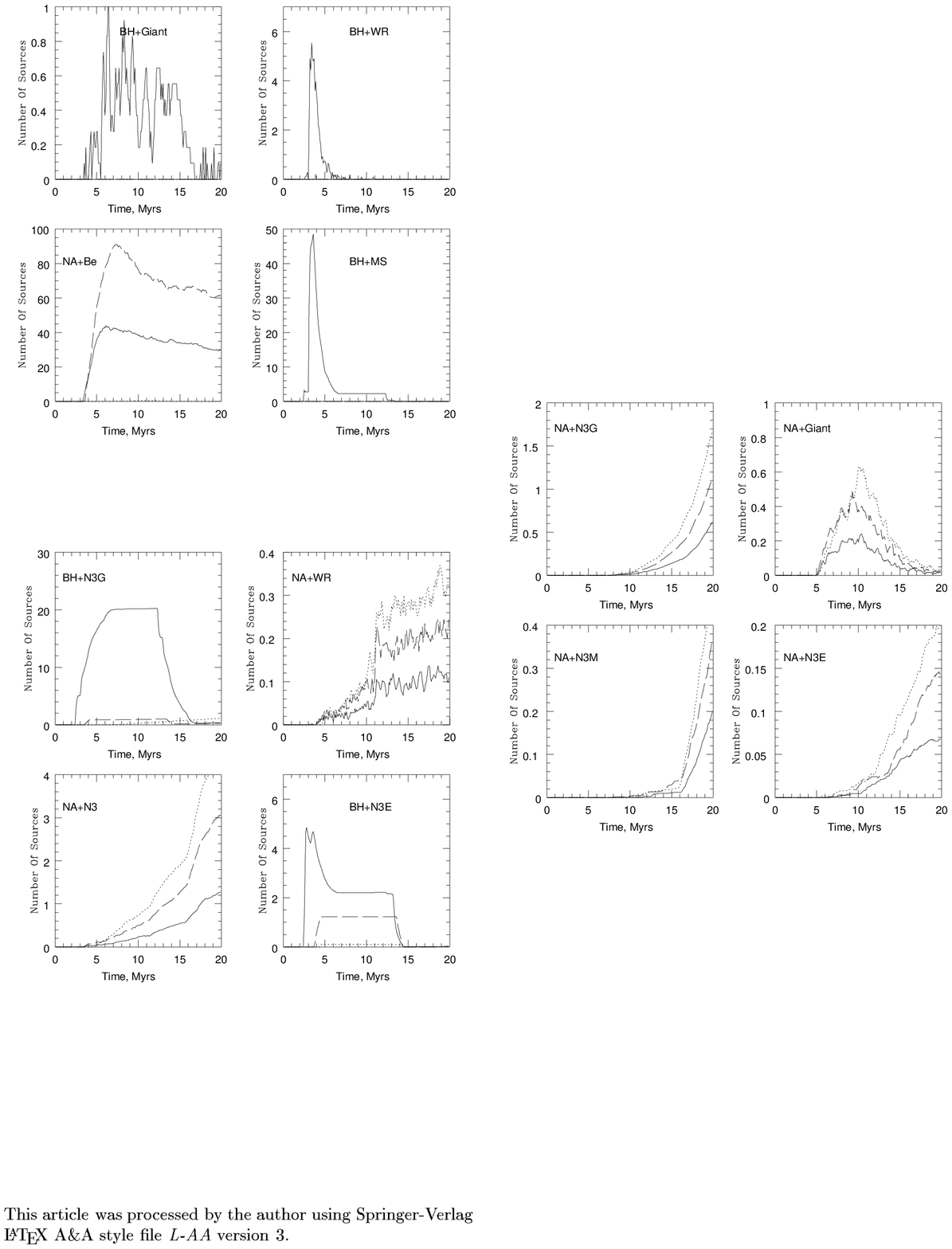}}}
\end{figure}

\begin{figure}
\epsfxsize=\hsize
\centerline{{\epsfbox{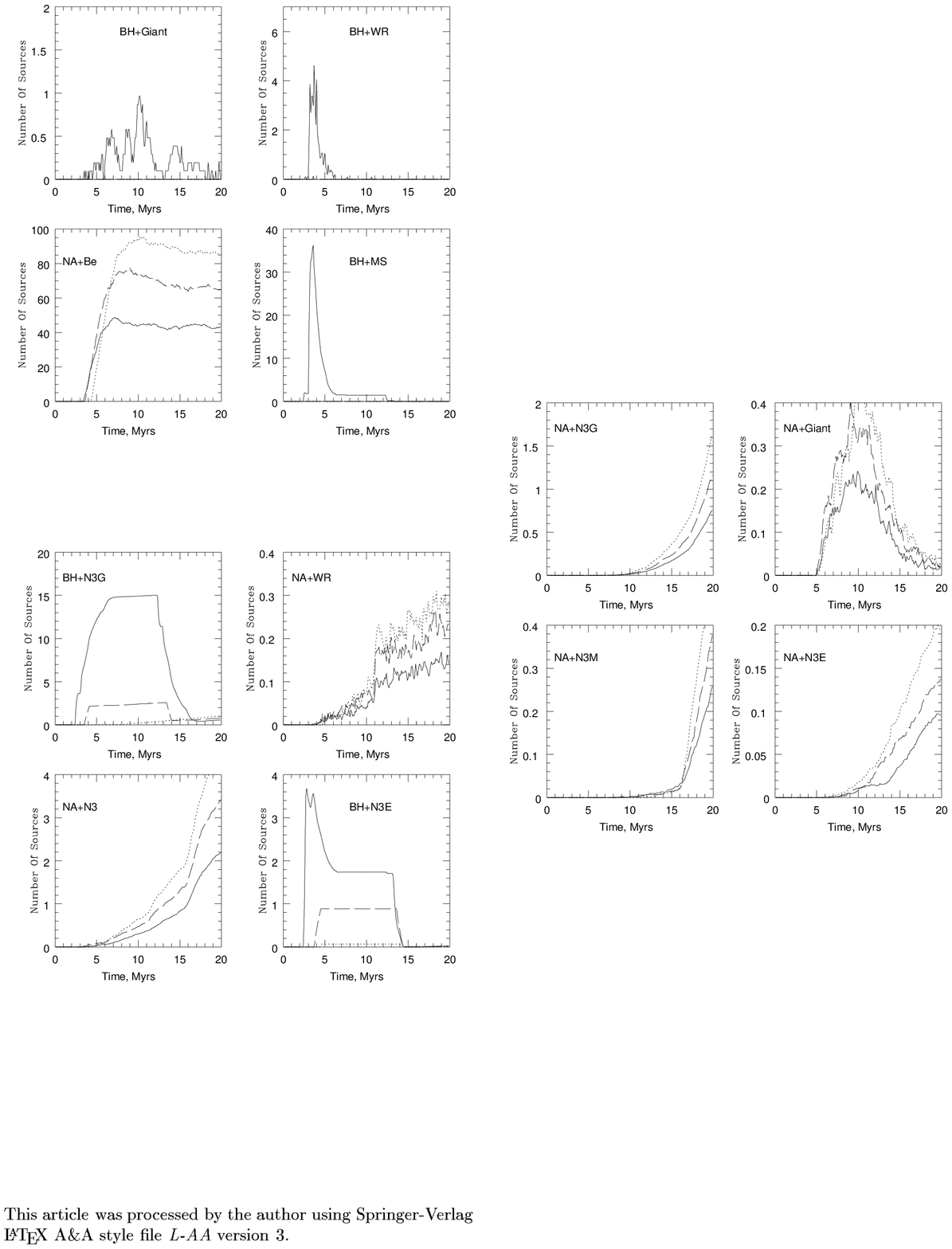}}}    
\end{figure}

\begin{figure}
\epsfxsize=\hsize
\centerline{{\epsfbox{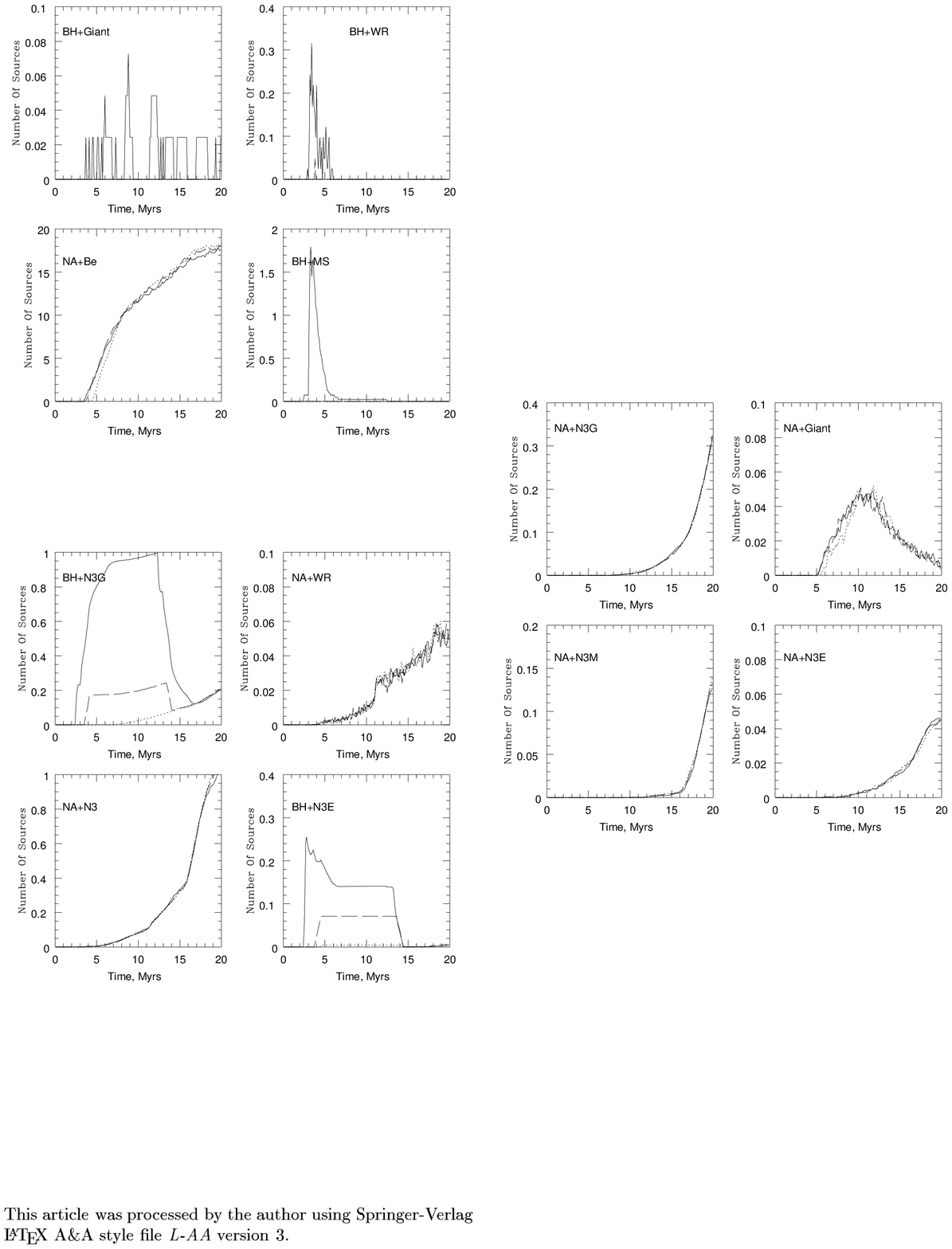}}}    
\end{figure}

\newpage
{\Large

\begin{center}

{\bf TWELVE TYPES OF X-RAY SOURCES}

\end{center}

\noindent
{\bf BH+N2} ---  A BH with a He-core Star

\bigskip

\noindent
{\bf NA+N1} ---  An Accreting NS with a Main Sequence Star 

\bigskip

\noindent
{\bf BH+WR} ---  A BH with a Wolf--Rayet Star
\bigskip

\noindent
{\bf BH+N1} ---   A BH with a Main Sequence Star
\bigskip

\noindent
{\bf BH+N3G} ---  A BH with a Roche-lobe filling star, when the
binary loses angular momentum by gravitational radiation
\bigskip

\noindent
{\bf NA+N3} --- An Accreting NSt with a Roche-lobe filling star
               (fast mass transfer from the more massive star)
\bigskip

\noindent
{\bf NA+WR} --- An Accreting NS with a Wolf--Rayet Star
\bigskip

\noindent
{\bf BH+N3E} --- A BH with a Roche-lobe filling star (nuclear
evolution time scale)
\bigskip

\noindent
{\bf NA+N3G} ---  An Accreting NS with a Roche-lobe filling
                  star, when the binary loses angular momentum due to 
                   gravitational radiation
\bigskip

\noindent
{\bf NA+N3M} ---  An Accreting NS with a Roche-lobe filling
                  star, when the binary loses angular momentum due to
magnetic wind
\bigskip

\noindent
{\bf NA+N2} ---  An Accreting NS with a He-core Star
\bigskip

\noindent
{\bf NA+N3E} --- An Accreting NS with a Roche-lobe filling star
(nuclear evolution time scale)

}


\section{Approximations. How to use it?}

 For the first set of our calculations we give analytical approximations
of our results.

 In the case of the Salpeter's mass- function ($\alpha = 2.35$)
and upper mass limit $M_{up}=120 M_{\odot}$ (see fig.1)
we have the following equations for X--ray transients
in the interval from 5 to 20 Myr after the burst (t-- time in Myrs):

\begin{equation}
    N(t)=-0.14\cdot t^2+5.47\cdot t -14.64   .
\end{equation}

 For superaccreting BH in the interval from 4 to 20 Myr:

\begin{equation}
    N(t)=\frac{2.2}{t-3.05}.
\end{equation}

 For Cyg X-1-- like sources in the interval from 4 to 20 Myr:

\begin{equation}
    N(t)=\frac{4.63}{t-2.9}  .
\end{equation}

   For binary systems with accreting NS and supergiants in the interval
from 5 to 20 Myr we have:

\begin{equation}
    N(t)=2.12\cdot 10^{-4} \cdot t^3 -9.6\cdot 10^{-3}
    \cdot t^2 +0.13\cdot t -0.47.
\end{equation}

    For ``flat'' mass- function ($\alpha=1$)
and upper mass limit $M_{up}=120 M_{\odot}$ (see fig.1)
for X--ray transients in the interval from 3 to 7 Myr we have:

\begin{equation}
    N(t)=-8.9\cdot t^2 +1.2\cdot 10^2 \cdot t -3 \cdot 10^2,
\end{equation}

\noindent
and in the interval from 7 to 20 Myr:

\begin{equation}
    N(t)=-2.8\cdot t +1.2\cdot 10^2.
\end{equation}

 For superaccreting BH in the interval from 4 to 20 Myr we have:

\begin{equation}
    N(t)=\frac{39.97}{t-3.17}       .
\end{equation}

For Cyg X-1 -- like sources in the interval from 4 to 20 Myr we have:

\begin{equation}
    N(t)=\frac{58.44}{t-3.08}         .
\end{equation}

   For binary systems with accreting NS and supergiants in the interval
from 5 to 20 Myr we have:

\begin{equation}
    N(t)=1.45\cdot 10^{-3}\cdot t^3 -5.96\cdot 10^{-2}\cdot t^2+
     0.74\cdot t -2.41.
\end{equation}

\newpage

\section{Discussion and conclusions. So what?}

Different types of close binaries show different sensitivity to variations
of the IMF. When we replace $\alpha=2.35$ by $\alpha=1$ the numbers
of all sources increase approximately by an order of magnitude. Systems
with BHs are more sensitive to such variations. 

When one try to vary the upper mass limit, another situation appear.
In some cases (especially for $\alpha=2.35$) systems with NSs show
little differences for different values of the upper mass limit, 
while systems with BHs become significantly  less (or more) abundant
for different upper masses.
Luckily, X-ray transients, which are the most numerous systems in our
calculations,
show remarkable sensitivity to variations of the upper mass limit.   
But of course due to their transient nature it is difficult to
use them to detect small variations in the IMF.

If it is possible to distinguish systems with BH, it is much better to use
them to test the IMF. 

The results of our calculations can be easily used to estimate 
the number of X- ray sources for different parameters of the IMF
if the total mass of stars is known.

In this poster we tried to show, that, as expected, 
populations of close binaries are very sensitive
to the variations of the IMF. One must be careful,
when trying to fit the observed data for single stars
with  variations of the IMF.

\end{document}